# THEORETICAL PHYSICS AND INDIAN PHILOSOPHY:
## CONCEPTUAL COHERENCE

Anna Sidorova-Biryukova
M.V. Lomonosov Moscow State University
119991, Russian Federation, Moscow, Leninskie gory, 1
E-mail: asidorova@mail.ru

> I have tried to read philosophers of all ages and have found many illuminating ideas but no steady progress toward deeper knowledge and understanding. Science, however, gives me the feeling of steady progress: I am convinced that theoretical physics is actual philosophy.
>
> Max Born

> ... with all our science, we are not a step closer to understanding the essence than an old Indian sage.
>
> K. Jaspers

## INTRODUCTION

These two epigraphs at first glance seem contradictory to each other; however, this contradiction could be better understood as complementarity, when two observers look at the same phenomenon from mutually orthogonal directions. One of the authors is a physicist, the other is a philosopher, and the subject they speak about can be called "metaphysics", or the science of eternal questions. The philosopher emphasizes their eternal character, i.e. the impossibility for a human mind to reach the finish line, while the physicist talks about the possibility and even the human duty to go along this endless path. This reminds of how Alice in the Looking-Glass Land had to run as fast as she could just to stay in place.

Indeed, as long as humanity remembers itself, it has been occupied by eternal questions. They changed their clothes, appearing in the form of religion, art, philosophy, and finally, science, but their essence remains the same. An Indian sage reflected on them, and modern science is trying to solve them.

As soon as we discover that two ideas from different disciplines make a "short circuit", we tend to believe that it is a sign of common essence that lies behind various manifestations. Although searching for analogies always comes with the risk of being too subjective and see just what one wants to see, it is still a powerful tool of cognition because points to some genuine values existing beyond here and now, which Karl Jaspers called "the cipher of transcendence" holding the "key to reality" [1].

Eternal questions are in the subject area of philosophy, but theoretical physics is another discipline that has to deal with them at a close range. It is no surprise then that the first study of parallels between two different worldviews --- Western rational and Eastern spiritual --- was the book of theoretical physicist F. Capra [2]. Published in 1975, the book soon became a bestseller, which was later translated into dozens of languages and has been successfully republished up to now.

In the 70s, when the book of Kapra was published, physics has been quickly and successfully developing, and most of scientists did not take seriously this sort of studies. The traditional Indian culture and philosophy were considered either as a collection of myths and legends of ethnographic value, or a set of unconventional techniques for body and mind, such as yoga and other occult items. However, it was not always so. Sometimes physicists found themselves in ideological deadlocks where they cannot longer hide from eternal questions behind the principle "shut up and count"; in this case, the most widely-thinking scientists turned to the wisdom of the East. It is well known that many founders of quantum theory were strongly addicted to Indian philosophy. For example, Erwin Schrödinger often referred to the Vedas and Upanishads, and Robert Oppenheimer knew Sanskrit and read Bhagavad Gita in the original, finding there a source of wisdom and inspiration. This interest of prominent theorists to the worldview of Indian metaphysicians was hardly mere amusement, it should be due to a deep correlation with their own representation of the reality [3].

For example, Robert Oppenheimer expressed this deja vu effect in 1954: "The general notions about human understanding... which are illustrated by the discoveries in atomic physics are not in the nature of things wholly unfamiliar, wholly unheard of, or even new. Even in our own culture they have a history, and the Buddhist and Hindu thought a more considerable and central place" [4]. An excellent selection of relevant statements of physicists, including the founders of the quantum theory Niels Bohr, Werner Heisenberg, Erwin Schrödinger, Wolfgang Pauli and many others is given in [5].

All this confirms the words of the philosopher and Hindu specialist Mircea Eliade, written in the middle of the 20th century: "The story of the discovery and interpretation of India for the European mind is really exciting" [6]. These words are no less relevant now. Today we observe a new rise of interest to the parallels; increasingly more authors notice resonances between modern ideas and the views of our far predecessors. A good deal of examples can be found in [7--14]. The proposed study is also intended to show that modern scientific theories are rooted deep in the experience of the previous generations of thinkers and to substatiate this fact by comparing some concepts of modern physics to the provisions of the Indian orthodox schools of philosophy.

Before we proceed to the main content, it should be determined what is meant by Indian philosophy here. As a rule, the ideas in question are taken from one or another of the so-called six orthodox schools of Indian philosophy. That is, from a specialist's view the comparison is made on a rather rough scale; however, we will adhere to the opinion of Max Muller, which he expressed in the conclusion of his monographical work "Six Systems of Indian Philosophy" [15]: "...it is wrong therefore to say of any of the admittedly orthodox systems of philosophy that it is not the means of right knowledge or that it is refuted by others. For in reality none of them is contradicted or refuted in what constitutes its own chief object." Further, he says that one can see "a kind of unity behind the variety of the various philosophical systems, each being regarded as a step towards the highest and final truth" (the principle of complementarity). And although Muller himself considers the systems "directly opposed to each other on very important points," he admits that their "followers managed to keep peace with each other and with the Veda, the highest authority in all matters religious, philosophical and moral." In addition, the original ideas can be distorted because much time has elapsed since then; therefore, "in trying to enter into the spirit of the Six Systems, we must implicitly trust to their guidance, without allowing ourselves to be disturbed by the fancies of later sects".

## 1. UNITY OF GOD, WORLD, AND HUMAN; KNOWLEDGE AS THE MAIN MISSION OF MIND

Holism is the fundamental principle of Indian philosophy. From a holistic position, the entire world is a single whole, while individual phenomena and objects make sense only as parts of it. The ultimate task of Vedanta (the basis of all the six classical schools) is to prove that the only true reality is the Supreme Spirit of Brahman and all the visible diversity of the world is the result of illusion (avidya), which should be abolished. Generally, the Hindu pantheism is well known; here it is interesting in view of its consonance with the so-called "theory of everything" [16] sought by physics today, whether it is quantum theory of gravity or another, but it should be a monoistic grand unification of all phenomena in a single description.

Striving for unity seems to be characteristic of our way of thinking and an immanent property of our nature, so this coincidence is quite expected, but another fact is surprising: from an enormous variety of religions and philosophies, only the Hindu schools pay so much attention to the metaphysical aspect that it even eclipses the traditional moral and ethical speculations, and try to make their theories as clear, consistent, and accurate as possible --- just in the same way as the science does. Max Müller admired by tireless and courageous Hindu thinkers, who had developed a "a system that even now makes us feel giddy, as in mounting the last steps of the swaying spire of an ancient Gothic cathedral. None of our philosophers, not excepting Heraclitus, Plato, Kant, or Hegel, has ventured to erect such a spire, never frightened by storms or lightnings" [15]. To ascend to the heights of Vedanta, one "must be able to breathe in the thinnest air, never discouraged even if snow and ice bar his access to the highest point". He further emphasizes the perfect character of this system, which is based on a strict and consistent logic: "Stone follows on stone in regular succession after once the first step has been made, after once it has been clearly seen that in the beginning there can have been but One, as there will be but One in the end, whether we call it Atman or Brahman" [15]. Indeed, studying the Hindu philosophy is felt much like that of a natural science: so strict is its structure, more suitable for a mathematical theory than for a religious philosophy. At the same time, it appears to be both abstract and mundane, closely related to the practical human life.

One the one hand, Hindu philosophy might be incriminated in promoting individual self-focusing and preferring personal intellectual work to the moral aspects of social life. Indeed, this is fairly strange for a religious philosophy, but is quite often encountered among theoretical physicists. However, by "lifting the Self above body and soul, after uniting heaven and earth, God and man, Brahman and Atman, these Vedanta philosophers have destroyed nothing in the life of the phenomenal beings who have to act and to fulfil their duties in this phenomenal world. On the contrary, they have shown that <...> goodness and virtue, faith and works, are necessary as a preparation, nay as a sine qud non, for the attainment of that highest knowledge". From the

Vedanta viewpoint, the ultimate human goal is to rejoin his consciousness with the Divine Cosmic Consciousness. There is only one way to reach this state of absolute freedom, that is, by means of congnition. Cognition is proclaimed to be the only effective remedy against suffering and boresom of mechanical life. (Nyaya even acknowledges the ability to cognize as the most immanent feature of a human.) This intellectual recipe of Hindu philosophy obviously differs from the solutions suggested by other philosophers of all the times. As Muller says: "None of them seems to me to have so completely realised what may be called the idea of the soul as the Phoenix, consumed by the fire of thought and rising from his own ashes, soaring towards regions which are more real than anything that can be called real in this life" [15]. Is there any other religion that sees the salvation of soul in sacrificing it on the altar of knowledge? This reminds us of the obsession and asceticism of great scholars, who lived in devotion to science often sacrificing the family and other common human joys. As for the moral code, it simply must be observed, as a necessary condition on the way to the spiritual summits; there is nothing to discuss for Hindu philosophers (although the practical methods that help people to live in accordance with this code are discussed, and in great detail).

What is this diverse and ever changing world from the viewpoint of a Hindu philosopher? It is no more real than the figure of clouds in the sky, which always change their shape and cause different associations with different observers. Physics today also describes the world as an eternally transforming system of more or less stable entities --- particles, fields. As the physicist Lorenz Krauss says, "we are all star dust," meaning that the atoms that make us now could be part of stars in other galaxies. However, behind ever-reassembling constructions of atoms and molecules there is something more real --- the patterns and laws of this assembly. Similarly, according to Vedanta, the visible world is like a mirage in the desert, but aquires reality in the Brahman. It is also important that avidya (illusion) is inseparable from human nature. What we perceive by means of our senses, as well as by means of various instruments, can never be absolute Brahman, but only His distorted image, which has passed through a nonperfect lens of our consciousness. Thus, Vedantists foresaw the problems that have led to instrumentalism but yet remained optimistic about our possibility to cognize the world, albeit by consequent approximations, building still more accurate theories, just like physics does.

In what follows we will note further similarities between Hinduism and science concerning the moral rating and the very spirit of these worldviews, and now let us outline more precisely the scope of ideas related to holism.

The unity of the world is the central idea, which is repeated in various forms by all the six schools. There are fairly straightforward comparisons:

"12-13 As one and the same string passes through gold, and pearls, jewels, corals, porcelain, and silver, thus is one and the same Self to be known as dwelling everywhere in cows, men, and in elephants, deers" [15, Chapter 6 ]. And also poetic metaphors:

"To view the bowels of the void deep all filled with thee <...> my peace dost fright." [17, Vol.II, p. 823]

Thus, the nature with its infinite diversity, hundreds and thousands of divine forms --- multicolored and diverse, is a product of the development of the One Reality. What is it? Modern science refers to this fundamental essence either as energy (if material), or patterns (if nonmaterial), or information (something intermediate). The prerequisites for each of these three concepts can be found in the Hindu philosophy. Let us start with the most obvious one, with patterns.

## 1.1. ONENESS AS PRINCIPLES

Thus, the kaleidoscope of visible phenomena and events in the world is an illusion, but there is no illusion without a reason, as there is no horror for the seeming snake without a rope lying on the road. Therefore, "having searched in their heart, found by wisdom the bond of what is in what is not" [15]. Modern physics considers the "bond of what is" as general principles (regularities, patterns) that are present in every phenomenon and can manifest themselves to an active researcher, who questions the nature by experiments and sometimes gets a hint to the truth. The following dialogue from Vedanta illustrates this idea of the invisible real present in the visible illusory.

1. The father said to his son: 'Place this salt in water, and then wait on me in the morning'. The son did as he was commanded.

The father said to him: 'Bring me the salt, which you placed in the water last night'.
The son, having looked for it, found it not, for, of course, it was melted.
2. The father said: 'Taste it from the surface of the water. How is it?'
The son replied: 'It is salt'.
'Taste it from the middle. How is it?'
The son replied: 'It is salt'.
<...>
The father said: 'Throw it away and then wait on me'.
He did so; but the salt continued to exist.
Then the father said: 'Here also, in this body, indeed, you do not perceive the True, my son; but there indeed it is'. [15, Chapter 4]

Every bit of the world embodies the general laws of nature, which means the omnipresence of the Supreme Spirit, striving for which is the highest duty and pleasure of the human mind. In recent time, we often speak about the principles in terms of information. This concept also has a strong resonance with the views of Vedanta, mostly in the context of virtual reality.

## 1.2. ONENESS AS INFORMATION

In the opinion of some modern scientists [18], information is capable of self-organizing, transforming the nature, and creating a new world, the so-called noosphere. For example, physicist David Deutsch defines information as an entity that supports, restores, and develops itself by organizing the environment [19]. (In case of biological organisms, this is quite obvious: the presence of a genotype in every biological individual is similar to the presence of spirit in an otherwise inanimate nature; but the definition also works for non-organic systems, for example, social institutions --- there is a good expression of "corporate spirit", which supports life within an association of people). This principle of running a software code by means of inanimate hardware can also be seen in the theory of the universe, as laid out by Paramahansa Yogananda. As he explains in the first chapter of the commentaries on Bhagavad-Gita [17], everything that happens in the universe is the Divine Play (lila) of the Supreme Spirit, who has divided Itself into numerous individual consciousnesses in order to "see the dreams of their personal beings". Each such being is nothing more than an information file, a plot arising in the Highest Consciousness. Then, human life is as illusory as the life of a computer character on the monitor and as individual as a wave on the sea surface; both are only examples of innumerable information processes on the surface of the Ocean of Omniscience.

Thus, a clear analogy can be traced between the virtual reality created by human consciousness and the reality of the world "created" by the Higher Consciousness (for more details see Section 5). It is interesting to note that this analogy might turn into identity (of the virtual and actual realities) if we recall of the principal engagement of the observer's consciousness into a quantum mechanical experiment. This would come very close to the Yogachara teachings, which asserts that the reality emerges as a result of the conscious activity.

The idea of cognition of the world by human mind as converting it into the virtual reality of knowledge has utmost expression in the omega-point theory of cosmologist Frank Tipler. In this theory, human mind facilitates the transition of the world to a state where "the universe will consist, literally, of intelligent thought-processes", and "the whole of space and its contents will be a computer" [19]. Let us compare this with the words of M. Eliade: "…when released, the man establishes the dimension of spiritual liberty and "enters" it into the Cosmos and Life, which are forms of conditioned existence and sadly blind" [6, Chap. II, Reintegration and Freedom]; the universal consciousness of enlightened creatures (like Krishna and Christ) is an "undistorted reflection of God permeating every atom and every point of space in the manifested Cosmos" [17, Vol.1, p.27]. The manifested cosmos is naturally the result of human activity. The similarity in understanding the mission of human consciousness in both cases is rather evident.

## 1.3. ONENESS AS ENERGY

In physics, the category of energy is actually introduced a priori, without explanation, as a "given by God". Indian sages frankly called this substance, from which everything is woven, cosmic dreams of the God: "... The yogi then rightly understands that <...> property of cosmic light is the building block of all objects and beings in God's dream cosmos" [17, Vol.2, p.679]. Here is the evidence of the universal character of energy, as well as an explicit mention of light, i.e. electromagnetic radiation, which is known to be the most elementary interaction in physics. Still closer resemblance to the modern concept of energy can be seen in "Samaniya" (common essence), which is one of the main categories used by Kanada, the author of Vaisheshika. Samania is assumed to be eternal, shared among many, and existing in the form of substance, property, or action. These three forms exactly repeat three forms of energy in physics: energy of rest, interaction, and motion. Samaniya can be high and low. Under certain conditions, high Samania is differentiated into several low types, which immediately resembles symmetry breaking and differentiation of interaction fields into the four types.

Let's summarize the first section.

In contrast to the popular belief about the "pantheon" of ancient Indian deities, it can be argued that the more important feature typical of the six orthodoxial schools of Indian philosophy is the concept of God as single and real essence of all the creative forces of the Universe; remarkably that this opinion is shared by some scholars of the XXth century [20]. (Regarding the "pantheon", Max Muller writes that "Kapila could hardly seriously believe in the Vedic gods (virgins), but he spares them, allows them to exist, perhaps bearing in mind that the people, honoring them, unconsciously approached the true purusha "[15].) Moreover, in Buddhism, in this limiting case of Hinduism, the concept of God as such is dissolved, and although "the goal is finding the one who acts, but the truth is that there is no one acting" [1].

However, we would not have talked here about the One Spirit if He could not be realized in some way. To translate into reality, He needed to imitate something different, second, dual, opposite to himself, but essentially one, namely, a man with his consciousness, where He could reflect in all its glory. So, the noble and difficult mission of the mind - knowledge and awareness of its unity with the Supreme Spirit transforms man himself, endowing him with infinite possibilities, and the Cosmos surrounding him, which completely turns into a conscious form of virtual reality (mathematical model of everything in the Universe). Both the followers of Vedanta and the representatives of modern science dream of fulfilling this mission.

The ambitious nature of this mission is combined with a democratic approach to the choice of its participants. As D. Deutsch writes: "If desired, any person can look for such evidence, find it and improve it. For this, no authority, no initiation, no sacred texts are needed. The only thing you need is to look in the right direction <...> This open accessibility is not only evidence, but also the whole mechanism of gaining knowledge "[19]. Similarly, for meditation nothing is needed except for one's own body and mind, and then, "all souls" will "return home, sooner or later [17, Vol.2, p.667]. Apparently, he and the other understood that even with the general mobilization of the population, very few would reach the heights of skill.

## 2. ENERGY, VIBRATION, LIGHT

Above we compared the Hindu fundamental substance (prakriti) with energy, but what does it particullaly look like? And again the answer of Hindu philosophers closely correlates with the present-day physics, which describes energy as vibrational process. Below we will show that waves, oscillations, vibrations, and radiation of light permeate Hindu metaphysics all through.

Through the practice of meditation, yogi learns to see the body not as a solid mass, but as a manifestation of vibrating light. He understands that the basis of physical cells is energy, which resides in the vibration modes of the Cosmic Mind. Vibrations of a body provide life in it, starting from the physiological processes up to refined mental activity. Patanjali defined thought as waves and ripples (vritti) in mind. He said that a "thought-wave rises, stays in the mind for a moment, then subsides, and the next wave follows it" [21]. Behind the poetic form, one can see a deep understanding of the fact that wave is nothing of a material but an information-carrying process.

Vibrations are the main type of work with body during meditation. Yogi learns to control the oscillatory processes of his body, starting with breathing and heartbeat and gradually moving to ever more subtle vibrations of the astral body and consciousness. The sounds of the Pandavas' battle shells in the first scene of Bhagavad-Gita can be interpreted as tuning to the Divine Vibrations, which become increasingly finer as one approaches the highest state of samadhi, attuned to the single Divine Wave of the universe. The same is the intention of the shortest and most popular prayer "Om" or "Aoum". One who pronounces it literally goes through the entire sound spectrum, from the maximum open "A" to the least that passes through the tight lips "M". Another example: vibrations on the sound "M", which are called brahmarie (bee's buzzing), are intended to revive every cell, fill it with energy, and prepare the body for meditation. Such an engineering or physiological approach to communication with God is characteristic of the Indian thinking and emphasizes its practical, body-oriented aspect, the necessity to verify all the techniques in one's own practice.

Returning to the forms in which the energy can exist, what is suggested by physics today? Actually, the very similar idea: to take a set of oscillation modes as a basis, be it particles of the Standard Model, strings, or some other fundamental elements. The stable existence of matter is also associated with cyclical motion: "a stable particle is a stable process, that is, a process that permanently repeats (or rather quasi-repeats) itself. In quantum theory, particles are endowed with an internal frequency that characterizes the cyclic changes of phase. This is postulated. In a discrete model, the presence of such a frequency is obviously a necessary property of any repeated structure" [22]. All the key concepts of the microworld theory are cyclical $\sim \exp(i\,\psi)$.

It is even more interesting that, along with the vibrating cosmos, the Hindu metaphysics admits also the existence of something nonvibrating; it is the sphere of the Absolute: "all human beings made in the image of God have a vibrationless, blissful soul within" [17, Vol.1, p.381]. From physical viewpoint, this category of "vibrationless" is attributed to something that shows invariance. An excellent example is the light, characterized with a constant speed of propagation. The fundamental nature of electromagnetic interactions is adopted by physics today. Especially intriguing is the interpretation of light as an inhabitant of the fifth dimension [23]. "Everything we see around us, from the trees and mountains to the stars themselves, are noting but vibrations in hyperspace" [24]. The main process that is observed in the Universe is the transform of the massive substance into radiation; in a sense, it is the return of matter into the Ocean of Light that was its origin.

It is also remarkable how the Nasadiya Hymn describes the beginning of times: "Darkness there was, in the beginning all this was a sea without light ; the germ that lay covered by the husk, that One was born by the power of heat (Tapas)" [15]. One can hardly stand from drawing an analogy between the "power of heat" and the Big Bang concept.

Light plays such an important role in this study that all the article can be devoted to describe it, but we will consider only one aspect related to the creative function of light as the fundamental principle of the world and, in fact, the only real substance from which it is woven. By fantastic coincidence, this idea is expressed by Johanada almost literally. We should remember that the image he found dates back to the mid-20th century, but hope that he conveys the original idea of the Hindu texts as accurately as possible.

"Just as the etheric flood of light going out of the movie booth is seen to be like a transparent searchlight free from any inherent pictures, yet images mysteriously appear on the screen; so God, from His booth in the center of eternity, is emiting a spherical bundle of rays <...>, which --- passing through the film of interacting principles of nature --- produce on the screen of space an endless variety of apparently real pictures" [17, Vol.1, p.123].

This actually holographic principle under the evolution of the universe that proposed by Indian metaphysicians has recently got its counterpart from the physics side: it was shown that events occurring in our 4D universe can be represented in terms of conditions and equations specified on its 3D boundaries (the concept of AdS/CFT correspondence) [25]. Thus, both the "film of principles" and the "spherical bundle of rays" (expansion of the universe in space-time) have got a scientific support. Science cannot yet answer whether the light is indeed the main creator of the universe, although some intriguing hypotheses appear [3]. Indeed, it is only the light that is massless, so it does not cause the space-time to curve, possibly, does not interact with it at all and therefore can exist beyond it. From mathematical viewpoint, electromagnetic interaction is also the most elementary since it obeys a commutative symmetry law and, hence, could exist "before" the moment of spontaneous symmetry breaking.

In addition to its key role in cosmology, the holographic principle should guide a person in the perception of his own body (which is a microcosm): "The yogi, peering with closed eyes into the dark invisibility within, finally finds there six subtle astral booths <…> He sees that the true-to-life picture of his body is produced by vibration in these centers" [17, Vol.1, p.123]. Various vibrations produce and support all the body tissues and activities. In this way the meditator comes to the deepest regions of the unconscious, which is experienced again as a bright flash of light. Modern neurophysiology suggests that these deep layers are the collective unconscious, supra-individual basis of the human psyche. This is another common feature of modern physics and meditative practice --- both of them explore areas that are beyond the scope of ordinary human perception and require sophisticated technique and years of hard training to be reached. However, there is an important difference: yogis perform all experiments exclusively on their own minds and bodies.

After so many arguments in favor of the fundamental nature of light, one may ask: perhaps all of its properties (constant speed, oscillatory nature, discrete character) are in fact not the properties of light, but of the system from which we observe it, that is, of our space-time?

### 3. SPACE AND TIME

With regard to the concepts of space and time, Hindu metaphysicists are rather unanimous considering them non-fundamental in many ways. Firstly, neither space nor time existed until the world appeared; as physicists would say now, they are of an emergent nature. In this point, the Nasadiya Hymn is similar to many other religious descriptions of the beginning of world: "There was then neither what is nor what is not, there was no sky, nor the heaven, which is beyond. What covered? Where was it, and in whose shelter?" [15, Chapter 2]. However, some nuances are noteworthy. In particular, the expression "neither what is nor what is not" emphasizes duality, i.e., anything that emerge should do it pairwise, where one of the two will belong to the visible world and the other --- to the hidden one. In physics, there are numerous manifestations of duality at various levels of the organization of matter, such as the particle-antiparticle pairs, left- and right-hand helicity of particles, supersymmetry, visible and dark matter, macroscopic and hidden dimensions, and many others. Furthermore, the question about the "shelter" under which the world was born bears evidence of the concern about the boundary conditions as an important factor that determines the development of the system.

Secondly, and this is also characteristic of Hindu philosophy, the ontological status of space and time is strongly associated with their perception by human consciousness: advanced yogi is able to say: "This world, this cosmos, are only shadows of life thrown on the screen of space, and reflected in our conscious and subconscious mental chambers" [15, I verse 1].

Buddhism even affirms the possibility for a person to exit from this conventional reference frame by means of a special yogic technique, intended for escape out of time. Buddha claimed he could very well exceed the time, using the "right moment", getting the "instant enlightenment", which "broke the Time "and enabled to get out of it "through a ruptured heavens" [6]. At the same time, the yogi "already lives not in time and not in subjection to time, but in the eternal present --- nunc stans --- as Boethius called eternity" [6, "The Techniques of Autonomy"]. Today's science is still trying to construct time machine, dig wormholes, or anyhow else to squeeze into the gap between worlds; although these attempts lie rather far from the mainstream direction.

It is interesting to note that the Hindu philosophers might link the relativity of phenomena with the illusory nature of space-time. On the one hand, they were well aware of the principle of relativity: "objects in the phenomenal world are called relative because they exist only in relation to each other. Man's ordinary consciousness is relativity consciousness — i.e., he apprehends one thing only by interpreting it relative to something else" [17, Vol.1, p.557], which is perfectly consistent with the tendency towards relationalism in physics nowadays. On the other hand, they considered any form of motion illusory except the oscillatory one. Indeed, consistently applied principle of relativity leads one to conclude on the equality of all observers that move in any manner in ordinary space. This eliminates all forms of motion, but not waves, which carry energy and, hence, represent something more fundamental than simply motion in space-time. Also, the carrying medium turns out to be crucially important: the vibrations of physical media (solid, liquid, gaseous, ethereal) occur in ordinary space, while the vibrations of "Supreme Spirit" (actually, the quanta of energy, as considered above) occur beyond the space-time. Therefore, the latter is only a screen where shadows of illusory movement are produced; this is precisely the conclusion that Hindu metaphysicists arrived at.

The illusion of motion in space-time is harmful because it masks the eternal and real; Samkhya explains how it occurs by the following comparison: an Individual Spirit (Purusha) is reflected in matter (Prakriti); and as a reflection in a mirror appears to be moving if the mirror moves although the reflecting object remains motionless, so the Purusha seems to be moving, although in reality it is only the Prakriti that moves.

Philosopers of Samkhya and Yoga viewed the categories of space and time as closely related, which is a good step towards the concept of united spacetime. Moreover, they admitted the existence of additional dimensions, which cannot be detected by our sensor perception. As Yogananda explained, there is a barrier between the four ordinary and higher dimensions: "beyond the subtlest physical vibration (as we would say now, the highest oscillation mode), there is the superether, a finer manifestation and therefore not classified as one of the physical vibratory elements <…> Some yoga treatises define this tattva as mind, or 'nonmatter,' as opposed to matter or gross vibration" [17, Vol.1, p.41]. The ideas of multidimensional world and strength fields as oscillations in hidden, compactified dimensions have been intensively studied by physics for more than a century, starting with the 5-dimensional Kaluza-Klein theory up to the string theories and even more exotic models like the "theory of everything" by Gerrett Lisi, which is built in eight-dimensional space [26].

## 4. EVOLUTION, REINCARNATION, CAUSALITY

The concept of evolution in the Indian philosophical schools differs essentially from what it usually means in the Western science. In Indian sense, evolution is not understood as becoming in time (the more so that time is rather an obscure cathegory); it is rather a timeless realization of ever-existing potential, which proceeds from elementary and fundamental to intricate and elaborated. As Eliade writes, "It is not a creation, not a surpassing, or realization of new forms of existence, but simply the actualization of the existing potentials in the prakriti". An attempt to identify the Indian concept of evolution to Western evolutionism leads to a "large confusion "[6, Chapter 1].

Here are some illuminating excerpts from the Samkhya Sutras [6, "Yoga Teachings"]: "From the moment he leaves his initial state of perfect balance and takes the specific conditions for its "theological instinct" the prakrti presents as a mass of energetic call Mahat (great)". Then, Mahat is divided into three gunas, which are types of energy, and each of them is manifested in the form of a particular class of phenomena. For example, the condensation of the tapas guna produces atoms and molecules, which in turn comprise plants and animals. This again confirms the unity of the universe, where everything originate from a single substance, be it atoms of matter or human consciousness. It is remarkable that the key moment in the emergence of "energy mass" is "leaving of the balance state," which looks much like a spontaneous symmetry breakage occurred to the initially uniform condition of prakriti.

It is noteworthy that this timeless complication of matter occurs spontaneously, actually for no reason. A parallel can be seen with the series of natural numbers: after making a few steps along this uniform sequence, one finds that prime numbers give place to nonprime ones, which have an increasingly complex internal structure; similarly, the initially homogeneous prakriti develops first into atoms and then into extremely intricate constructions like human consciousness. This nontrivial idea of the internal dynamics embedded in the very structure of an unfolding evolution process demonstrates that the thinkers of the past had got a very deep insight in what can be called the "permanent creation of the world" [3].

Sometimes, the subsequent emergence of events in Samsara (Buddhism) is compared to the reflection of one source of light in two mirrors posed in front of each other. So arranged mirrors produce an infinite number of reflections, and this infinity gives rise to a new quality: thus, the reflection of natural numbers from zero point gives rise to integers, reflection of integers from the fraction bar gives rational numbers, then in the similar manner real and complex numbers can be obtained. Of particular interest is the question whether this process can generate a continuum, for example, the space-time continuum starting from discrete elements. It has been shown recently [27] that such process is free from logical contradictions.

The magic of reflection has been noted at all times; mirror images were used by Buddha and Dante, Paul Florensky [28], and Lewis Carroll. Indeed, mirroring of a system should necessarily give rise to new elements and new properties, since there are no means to obtain the result of reflection while staying within the original framework. In a sense, this process is analogous to the known Cantor's diagonal method of constructing a qualitatively new theorem, which can be neither proven nor disproven by means of the initial axiomatic set. On a worldview scale, the "right" to "left" change can be regarded as a transition from the internal space of a system to its external environment, where different laws and new opportunities may encountered. In Indian mythology, a higher world of Gods is sometimes interpreted as a mirror image of the world of people. Later, this idea has received highly artistic embodiments, e.g. in the Carroll's novel "Through the Looking Glass" and Strugatsky's "The Doomed City".

There are still plenty of small details in Indian metaphysics that sound up-to-date now. For example, Vaisheshika teaching supposes that atoms "form first an aggregate of two, then an aggregate of three double atoms, then of four triple atoms, and so on. While single atoms are indestructible, composite atoms are by their very nature liable to decomposition, and in that sense, to destruction" [15, Chapter 9]. Concerning the breakage of initial symmetry, amazing details can be found in Bhagavad-Gita: "... God's consciousness which when differentiated becomes eight intelligences, or "eight sons": the Universal Unchangeable Spirit shining everywhere in the universe; six Spirits governing the three macrocosmic manifestations and the three microcosmic manifestations and the reflected Spirit (from the Universal Spirit)" [17, Vol.1, p.36]. In addition to the idea of lost balance as a mechanism of the world creation, we find an intriguing characterization of the Son Spirits: one stands separately in a privileged position, three others have macroscopic embodiment --- this 1+3 structure clearly reminds us of the space-time signature; while the reflected "microscopic manifestations" can be compared to the compactified dimensions of the momentum space in which quantum mechanics describes microscopic events.

(*Footnote) Paul Florensky suggested that the area of a 2-dim figure changes sign when viewed "from the back side", or being mirrored; therefore, the imaginary values of lengths can be interpreted as a reflection of real ones, while their pairwise combinations give complex numbers. (*)

It is noteworthy that the Tippler's Omega Theory (mentioned above) considers the ultimate limit named omega-point as the transcendence of physical reality into its virtual copy; at this point, the entire universe will be transformed into information loaded into a supercomputer and re-created in the form of virtual reality. However, the realization of this great project requires the last step to be made in the chain of Prakriti complication: One Spirit → space-time → massive substance → organic life → human consciousness → collective consciousness. It is known that, "when solving global problems, humanity acts as a single organism" [29]. It is straightforward then to suggest that the next step in our evolution will unify human intelligence on the global scale (for example, by means of the Internet). Possibly, the unified planetary brain will become one of the structural elements of the "Cosmic Mind", its single atom, which will comprise molecules and larger systems that will be able to realize the Divine Purpose.

Speaking about the concept of evolution in Indian philosophy, it is impossible to dismiss the idea of multiple reincarnation of a soul in different bodies. In essence, reincarnation can be seen as the life of numerous copies of one and the same information code, a genome, placed into different conditions and having various destinies. This idea actually anticipates the Everett multiverse concept (given in this context, e.g., in [19]) and its mathematical basis --- the Feynman principle of path summation over all possible trajectories. The task of every living soul is to make its path through the life in an optimal way; however, each next step along this path is not strictly determined but spread into a spectrum of more or less likely variants; hence, only taking into account the entire spectrum of possibilities gives all the necessary information about the path and allows the wandering soul to determine the right way.

Finally, the timeless nature of evolution entails that causality also has no direct connection to time. According to Nagarjuna, when the means of knowledge at least succeed in knowing something, they are no more means of knowledge without reference to the known objects. Therefore, there is interdependence between the means and objects of knowledge. He compares these relations with relations between father and son. "A son <...> dependent on a father, and a father is impossible without a son. In the same way everything is dependent on something else" [15]. Who is the father, and who is the son, who is born by whom, asks Nagarjuna, and how to distinguish between the means and objects of knowledge if, being establishers, they are the means of knowledge and, being established, the objects of knowledge. The situation fits perfectly into the conceptual scheme of binary geometrophysics, where "instead of the causality principle, a universal connection between all possible phenomena is postulated" [3].

The same timeless connection between events is seen in the canonical example of a statue, which has always existed in stone, including the time before the sculptor made it visible. In the same manner, there is no source without a receiver: "As there is no hearing, without sound, the Samkhya seems to have argued, neither is there any sound without hearing" [15, Chapter 6]. Similarly, in the modern world (much of which has already turned into the virtual reality), a picture posted on the Internet literally does not exist until someone wants to look at it and bring it to life by a keypress.

This scope of ideas includes also the concept of action at distance and a new understanding of locality. Swami Sri Yukteshwar wrote in Holy Science: "To keep company with the guru is not only to be in his physical presence (as this is sometimes impossible), but mainly means to keep him in our hearts and to be one with him in principle and to attune ourselves with him". The Internet has realized this new kind of locality: years ago people had to interact mainly with their neighbors in physical space, but now this restriction is removed owing to the global network, everyone can communicate/interact with anyone and the notion of locality as close position in physical space is replaced by that in the space of interests and views. It is this understanding of locality as proximity in the space of visions that is used in the model of real quantum ensemble in physics [30].

## 5. CONSCIOUSNESS AND ENTANGLEMENT

So, the world is woven from an ever-existing and permanently varying substance. Then, why we perceive it as something given here and now and all our personal experience, based on sensory feeling, is firmly connected with the sensation of a fixed present moment? To solve this paradox, Samkhya philosophers assumed that prakriti remains active only until it is noticed or perceived by Individual Spirit, Purusha. Under the glance of Purusha, Prakriti turns into buddhi --- conscious perception; the interaction with mind makes the substance both perceiving and perceived.

It is easy to see here an analogy with the wave function collapse in a quantum experiment. Philosopher of science William James said that the true result of an experiment is the intellectual interpretation of its data. According to modern views, it is impossible to divide the content of a quantum experiment into subject and object. As N. Moiseev notes, "even the acquired knowledge and the picture of world arising in the minds of scientists produce an influence on the evolution of the world that we live in!" [31] Owing to the entangled observer, the experiment becomes creative and creates that same reality that we detect in this experiment. At this point, the mysticism of Upanishads is in perfect agreement with quantum theory, where the world and consciousness are entangled into an unseparable whole. "That are Thou!" says the Upanishads, and the Vedantists add: "Nor a part, nor a mode of That, but identically That, that absolute Spirit of the World" [32].

This self-consistent concept is another evidence of how causality is replaced by mutually equal relation. Just as Brahman creates the world and human consciousness by his cognition, so Atman, which is essentially the same as Brahman, by his cognition creates Brahman as his creator. The connection of Brahman and Atman is the relation beyond time and space, similar to that in the previous examples of the relations between father and son, source and receiver.

If the reality and consciousness are so closely entangled, then the irreversible collapse of state in a quantum measurement can be considered as a phenomenon of consciousness. This viewpoint is suggested, for example, by M.B. Menskii. The concept he develops is highly consonant with Vedic mysticism: "The world is only an image arising in consciousness. If consciousness percieves only one of the Everett worlds, then the other worlds do not seize to exist <...> The idea that only one world chosen by consciousness is real is just an illusion that arises in the mind of an observer" [33]. In the quantum world, there is no reduction of the wave function and, hence, no difference between the present, past, and future, and also between the Everett worlds.

If mind could penetrate into the multiverse it would get quite a new level of knowledge, including information from any time and any "parallel reality". Whether it is the state of meditation, which is known, for example, in yoga practices? Indeed, each moment we perceive only one reality, one world cut out of the multiverse. Perhaps this is because we are fated to move always in one direction along the time axis. Meanwhile, if one would learn to ignore this "cutting out" of a single reality and come into a state of silenced consciousness, he could feel the other dimensions of the world and obtain unusual new opportunities. Meditative techniques allowed oriental sages to discover great mysteries of life, and the fighters who mastered this skill were invincible, because they knew every move of the opponent before he even thought of doing it [34].

A few years ago, an intriguing study in neurophysiology appeared, which is directly related to this discussion. In her innovative work, neurophysiologist A. Sverdlik argues against the conventional opinion that it is the young cortical sections of brain that are mostly responsible for abstract mathematical thinking; instead, she asserts, mathematical problems are solved mainly by means of the ancient deep sections of brain. Moreover, the more complicated and abstract the problem, the deeper layers of brain involved in the search for solution. The somatic nervous system only "splits the world around us into colors, smells, and sounds, into objects, objects, and objects again <...> This is the very fog into which it immerses our abstract thinking: the essence of things is always obscured by a dense veil of fragments --- small parts of the world around us" [35]. To continue this idea, one can suggest that for solving hard problems we can use not only the brain, but involve the whole body as well. This principle is the core of the yoga techniques, where the work with body and breath helps one to awaken the ancient layers of the primary consciousness. Many scientists noted that solution to the problem on which they were working for years came to them unexpectedly, at the moment when the consciousness was close to meditative state. In the opinion of R. Penrose, it is due to our affiliation to the universe and the opportunity to exercise physical laws in our own bodies that human mind can solve problems unsolvable for artificial intelligence [36]. This substantiates the presence of a deep, physiological level of parallels between mystics and physicists. Generally speaking, there is no surprise here, since both deal with close subjects, as E. Witten said: "…physics is about concepts, the principles by which the world works" [24] --- just the same cipher of transcendence in terms of K. Jaspers.

## 6. INVARIANCE AS REMOVAL OF PROPERTIES

As M. Eliade noted, meditation is not a detachment from reality, immersion in dreams and hallucinations; on the contrary, it is a rise to new levels of reality, a break through the veil of momentary circumstances and properties. Katha Upanishads say: "13. A mortal who <…> has removed from it all qualities, and has thus reached that subtle Being" [15]. But what is mathematics if not the same elimination of properties? Indeed, it builds more and more abstract theories, starting from arithmetics, which uses the same natural numbers to count various objects, to ever higher levels of abstractness, thus, embracing and explaining an ever wider scope of phenomena.

The process of "removing properties" refers to apophatic methods of cognition [37] and finds the ultimate expression in Buddhism, which teaches that nothing can be asserted for sure. Since we obtain properties as a result of measurement or observation, hence they characterize only a projection of the phenomenon on the reference system of the observer; this means that different observers will get different values of the properties, which are all turn out to be fiction then. This situation reminds the story of the seven blind who grope an elephant. But there is another Indian parable about a lame man who sat on the shoulders of a blind man, which was to the benefit of the both. This parable shows that the Indians understood perfectly well the principle of complementarity and how mutually exclusive viewpoints can all be true. The Indian tolerance to alien opinion is well known, but not so is the fact that they invented a system of many-valued logic (Syadvada) [38], where any statement is considered possible with some probability; here one can easily guess a forerunner of probabilistic values in the quantum theory.

Max Born also talks about the elimination of properties in "Physical Reality" [39]: in his opinion, the most expressive fact of our spiritual structure is the ability of our soul to ignore variations in sensory feelings and to note only their invariant components. The measurement itself is always a positive experience, but as soon as mind switches on to analyze these data and make an inference, it is obliged to use negation. This happens because the boundary between the measurement tool and the measured object is obscure; any physical experiment determines the value with only a finite accuracy, that is, it only cuts off the range where it cannot be found. However, this uncertainty does not preclude the cognition of reality. As a result of additional experiments, there arises a group of invariants that characterize the entity under study [39]; for example, the invariant essence of a 3-dim elephant can be reconstructed in three mutually orthogonal projections, each of which actually defines the area of space where there is no elephant. In scientific research, the more such additive negation experiments, the more accurately we approach the invariant essence. And in this sense, we can say that "inconceivable is conceived by inconceivability" [37].

It is known that in Buddhism the absence of an answer was not necessarily attributed to ignorance, but considered as one of possible answers; this is a common situation in mathematics. This distinction between not-knowing and the absence of a solution has become especially clear after the advent of artificial intelligence; if a problem has no solution, human mind is able to guess this fact and, thus, to solve the problem, while computer will just keep cycling forever.

## 7. TRINITY, TETRADITY, SYMMETRY

Invariance is a form of symmetry. Probably, there is no any other religious or philosophical system that worships symmetry as much as Hinduism. Suffice it to recall the highly symmetrical geometric patterns of yantras and mandalas. The basic form of most mandalas is a square with four gates containing a circle with a central point [40]. The cult of perfection is also traditional for the Hindu culture: it can be seen either in the requirement of doing yoga asanas with geometrical precision or in the strict adherence to ritual procedures. Any action made perfectly is attributed to have magical force. Possibly, this worship of perfect symmetry may seem too naïve and straightforward, but it agrees remarkably with the scientific position of physics. As S. Weinberg says, "the principles of symmetry and different modes of wave function transformations using these principles rather than matter itself dominate in physics" [16].

Concerning the particular details of symmetry, they are not so important as the main principle itself; however, it is interesting to note that, unlike Christianity, Induism prefers even numbers, especially four and eight, rather than three. The adherence to parity corresponds to the reign of duality in the Hindu dialectic. As for the triad, first Western researchers of India tried to draw a parallel between the Christian Trinity and the Hindu Trimurti, which is the trinity formed by Brahman, Vishnu, and Shiva. In fact, this similarity is deceptive; as V. Shokhin says, "from the scientific viewpoint of indology, the comparison reveals only the fundamental differences between these two concepts" [41]. Trimurti was, in fact, an artificial value and was never popular, unlike the Holy Trinity.

Rather than in the Hinduism religion, Triadism manifests itself more naturally in the Indian metaphysics, where the primal state of prakriti is considered as a state of balance between three gunas: sattva, rajas, and tamas, while all events in the Universe occur due to the loss of this balance. On the human scale, the three gunas are displayed in three human bodies: robust physical, sensual astral, and subtle spiritual. Concerning graphical manifestations, triada appears mainly within a composition of six-fold figures, where its imperfect oddness is somewhat damped by multiplication by a factor of two.

The most right number in Indian metaphysicists is definitely the number of four and of the doubled four, eight. The geometric symmetry of a square lies in the heart of Sanskrit (the "perfect language") and has numerous manifestations in Indian metaphysics, literature, and spiritual practices. Physicists like square since it is the form of a two-dimensional Cartesian coordinate system, which is still the most convenient way to visualize physical graphs and dependences.

## 8. PRACTICAL, PEDAGOGICAL, ETHICAL

As was stated above (Section 6), pure objective knowledge can be obtained by depriving the perceived information of the subjective empirical component. But who can perform such a cleansing better than the knower himself? Here is another fundamental point of the Indian philosophy --- one can climb to its summits only via an honest and permanent self-working, doing one's personal practice.

At the same time, it looks rather unusual that this advance towards spiritual heights does not mean removing from every-day problems; on the contrary, the highest wisdom is intended to effect daily activities, making them more effective. This is another aspect of oneness: the overall harmony should govern concrete practical deeds. Moreover, it is the practical applicability of knowledge that makes the best criterion of its truth.

The practical orientation of Buddhist and Indian philosophic teachings is often noted and even opposed to those of Western philosophies, which usually experience a gap between thought and action. Ideas in Indian metaphysics should be verified by practice, and in this point it is close to modern science, where experiment is the prior criterion of truth. A pupil should not take what guru says on trust but try to check as much as possible in his own practice and, thus, reduce the number of unreasonable axioms. Physics also tends to minimize the number of a priori input assumptions of a theory and tries to explain as many phenomena as possible using as few premises as possible.

It is interesting that, in the opinion of Vaisheshika adepts, the sacred books should not speak about anything beyond the sphere of knowledge of ordinary people, for example, about reward for sacrifice in another world and other things beyond human experience. Indeed, the content of some texts (for example, the sutras of Patanjali) is so concrete that resembles a learning course for working with body and mind; while the simple and clear presentation in the form of aphorisms calls for the reader to start doing it immediately.

Probably, no other religion but Hinduism is so decisive in choosing reason out of the "reason or piety" alternative. Comprehended study and research are unequivocally encouraged as the main purpose of human beings: "Blind piety is not unacceptable to the Supreme Being, but is a low form of spiritual mindedness" [17, Vol.1, p.135]. God is delighted to see his human children using His highest gift --- their divine innate right of intellect.

Not only the intellectual orientation of Indian philosophy, but also its ethical code can impress the modern scientific community. "Indian philosophers are truthful, and Patanjali says in so many words that truth is better than sacrifice. They may err, as Plato has erred and even Kant, but they are not decepti deceptores, they do not deceive or persuade themselves, nor do they try to deceive others" [15].

A compliment to Buddhism sounds in the words of Nietzsche that "Buddhism is one hundred times colder, more truthful, more realistic and more objective than Christianity". In addition, such features of Hinduism as the rejection of coercion and tolerance for dissent, refusal of austerity but healthy moderation in material needs, fighting with human suffering rather than human sins, the desire to know God instead of the fear of Him --- all these can only be welcomed by modern science.

## CONCLUSION

For centuries, representatives of European civilization unwillingly experienced some sense of superiority in relation to other peoples, which they subordinated (often, by force) and by this reason considered to be "lower" or even "primitive". It took "the thrust of European metaphysical thought-border of this century, the religious revival, the multiple procurement action of depth psychology, poetry, the microphysics, in order to understand the spiritual horizon of the 'primitiveing', the structuring of its symbols, the function of its myths, its mystical maturity" [6]. Although it is believed that the technical progress has led our civilization to the top of human development, in fact it is only getting to the same inferences that the Oriental thinkers made several thousand years ago, and without using 'instruments', but only with the help of one's body and mind.

The harmonious and logically perfect system of Indian philosophy most of all resembles a network, where all concepts are firmly connected with each other into a single whole. This is much of a system that physics strives to construct today. Since the time of Newton and until now, science was describing the world independently from human consciousness, which was considered to be a bystanding observer. Ultimately, this Cartesian division of world into human mind (congniting subject) and nature (cognited object) promoted the consumer attitude of human society towards nature and stimulated the utilitarian component of science. Now, when it turned out that human consciousness should also be included into the network of universal relations, the recreation of pre-Cartesian unity may return science its original metaphysical orientation and make each person's life "profoundly meaningful" by this "experience of belonging to the universe" [42].

"The best about what it is to be human — our ability to imagine the vast possibilities of existence and the adventurousness to bravely explore them — without passing the buck to a vague creative force or to a creator who is, by definition, forever unfathomable" [43]. These words of physicist L. Krauss are remarkably consonant with the very spirit of Indian classical philosophy, which inspires and supports human mind in its active and conscious effort to cognize nature, thus, gaining wisdom and freedom.

Speaking about the beneficial influence of philosophy on science, one cannot ignore the reverse process, as V. Vernadsky said: "The growth of science inevitably leads to an extraordinary expansion of the sphere of philosophical and religious awareness of human spirit; by perceiving scientific achievements, religion and philosophy further expand the deep recesses of human consciousness" [44]. Probably, in this sense we should also understand the words of M. Born in the epigraph that "theoretical physics is the true philosophy".